\documentclass{mem}
\usepackage{natbib}\usepackage{txfonts}\usepackage{balance}
\usepackage{graphicx}
\usepackage[a4paper]{hyperref}
\idline{000}{000}
\begin{document}
\def\teff{$T\rm_{eff }$}
\def\kms{$\mathrm {km s}^{-1}$}
\newcommand{\ltappeq}{\raisebox{-0.6ex}{$\,\stackrel
        {\raisebox{-.2ex}{$\textstyle <$}}{\sim}\,$}}

\title{Cataclysmic Variables in Globular Clusters}

\subtitle{}

\author{Christian Knigge}

\institute{
Physics and Astronomy
University of Southampton
Highfield
Southampton SO17 1BJ
United Kindgom
\email{C.Knigge@soton.ac.uk}
}

\authorrunning{Knigge}

\titlerunning{CVs in GCs}

\abstract{
Every massive globular cluster (GC) is expected to harbour a
significant population of cataclysmic variables (CVs). In this
review, I first explain why GC CVs matter astrophysically, how many and
what types are theoretically predicted to exist and what observational
tools we can use to discover, confirm and study them. I then take
a look at how theoretical predictions and observed samples actually
stack up to date. In the process, I also reconsider the evidence for
two widely held ideas about CVs in GCs: (i) that there must be many
fewer {\em dwarf novae} than expected; (ii) that the incidence of
magnetic CVs is much higher in GCs than in the Galactic field.}


\maketitle{}

\section{Introduction}

Globular clusters (GCs) are old, gravitationally bound stellar systems
that typically contain $\sim 10^6$ stars. Some of these stars are
binaries, and some of these binaries are cataclysmic variables (CVs),
i.e. systems in which a white dwarf (WD) accretes material from a
roughly main sequence (MS) companion. These CV populations in GCs
deserve special attention for at least three reasons: 

{\em 1 The Globular Cluster Perspective:} the late dynamical evolution of
a GC is thought to be driven largely by its close binary (CB) population 
(e.g. Hut et al. 1992). However, the dominant non-interacting MS-MS
binaries are difficult to detect and study in GCs. CVs can therefore
be used as convenient tracers of the underlying CB populations for
studies of GC dynamics and evolution. 

{\em 2 The Cataclysmic Variable Perspective:} in principle, GCs can
provide us with sizeable samples of CVs at known distance and (to some
extent) age. Such samples might allow critical tests of theoretical
CV/binary evolution scenarios. An interesting complication is that not
all CVs in GCs are likely to have formed and evolved in isolation: 
many are likely to have been produced, or at least affected, by
dynamical interactions (see later). 

{\em 3 The Supernova Perspective:} it has been suggested that GCs
might be significant Type Ia Supernova factories (Shara \& Hurley
2002), at least in elliptical galaxies (Ivanova et al. 2006). All SN
Ia progenitor populations are thought to be
close relatives of CVs (e.g. double WD systems, supersoft sources, WD
+ red giant binaries), so CVs can again serve as a useful tracer 
population for these progenitors. In fact, it is even still possible 
that some CVs might be SN Ia progenitors themselves 
(Thoroughgood et al. 2001; Zorotovic, Schreiber \& G\"{a}nsicke 2011).

\section{Theoretical Background}

So how many -- and what type of -- CVs might we expect in a typical
massive GC? Let us start with a simple back of the envelope
calculation, based solely on the number of stars in a cluster and
entirely ignoring stellar dynamics. The space density in the Galactic
field is of order $\rho \sim 10^{-5} {\rm pc}^-3$ (e.g. Pretorius et
al. 2007; Pretorius \& Knigge 2011). The volume of the Milky Way is of
order $V \sim 10^{11} {\rm pc}^3$, so the expected number of CVs in
the Galaxy is around $N_{CV,MW} \sim 10^{7}$. Now the fraction of the
Milky Way's stellar mass that is bound up in its GC system is roughly
$f_{GC} \sim 0.001$ and there are $N_{GC} \sim 100$ GCs in the
Galaxy. So, {\em other things being equal}, the number of CVs that
would be expected in a single GC is of order $N_{CV,GC} \sim f_{GC}
N_{CV,MW} / N_{GC} \sim 100$.

\begin{figure*}[t]
\label{ivanova}
\begin{center}
\resizebox{0.9\hsize}{!}{\includegraphics[clip=true]{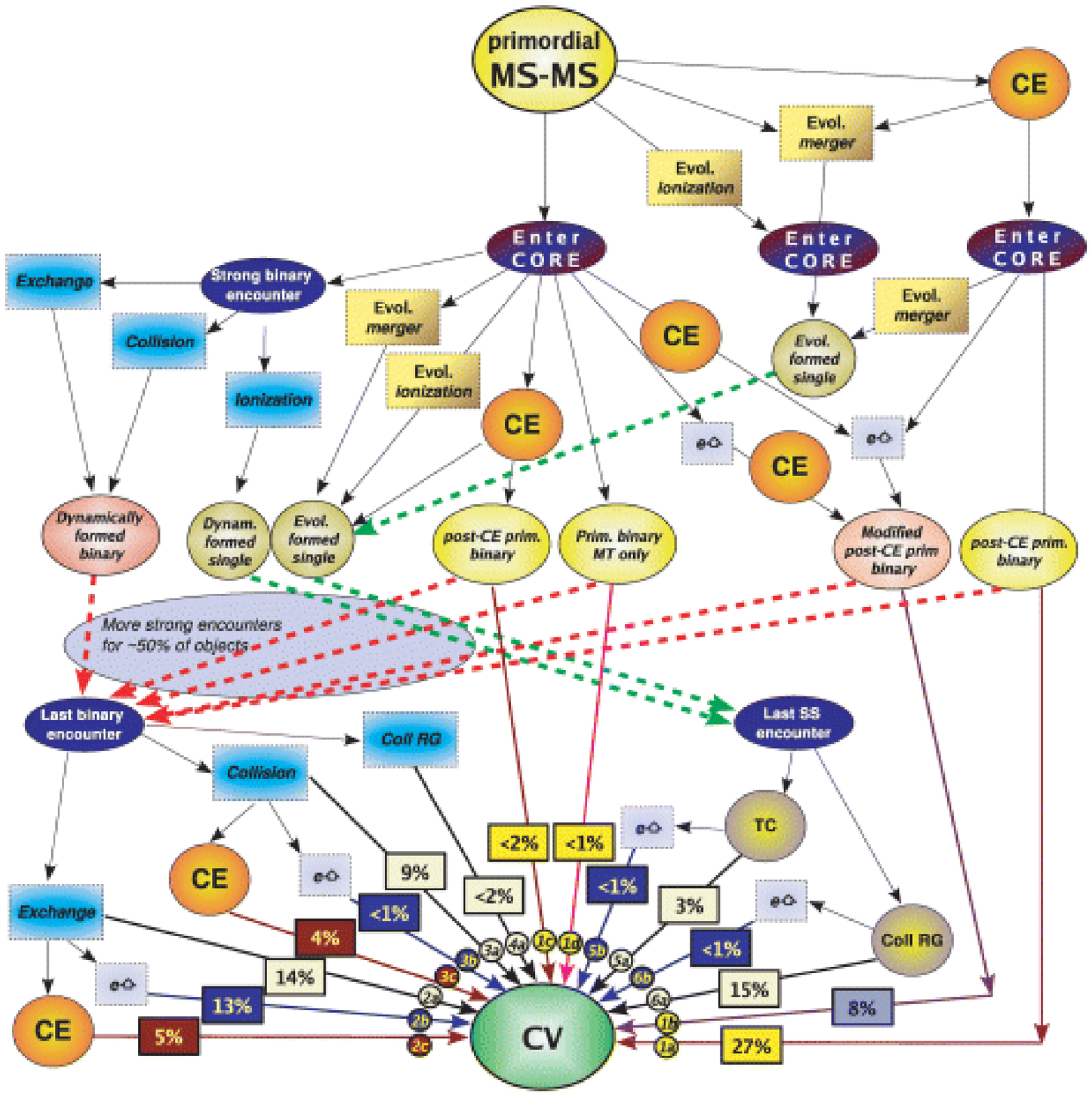}}
\end{center}
\caption{
\footnotesize
The main CV formation channels in GCs in the simulation carried out by 
Ivanova et al. (2006), to which the reader is referred for
details. Figure reproduced by permission from Ivanova et al. (2006).
}
\end{figure*}

But {\em are} other things actually equal? We already noted above that
CV populations in GCs might be strongly affected by dynamical
encounters. This is actually quite easy to understand: stellar
densities in dense cluster cores can reach $10^{6} {\rm pc}^{-3}$. To
put things into perspective: the volume occupied by a single nova
shell, like that around T Pyx, is $\sim 0.01 {\rm pc}^{3}$. So in a
dense GC core, this same volume will contain $\sim 10,000$
stars. Close dynamical encounters between cluster members are
therefore inevitable.

Fundamentally, there are two types of relevant dynamical encounters:
direct collisions and near misses. These can affect CV populations in
GCs in three basic ways: (i) they can create new CVs; (ii) they can
destroy wide binaries that would have evolved into CVs in the field;
(iii) they might alter the binary properties of CVs/CV
progenitors. It has been known for a long time that bright,
neutron-star-hosting low-mass X-ray binaries are $\simeq 100\times$
overabundant in GCs, presumably because they are efficiently produced
by dynamical formation channels in the dense GC environment
(e.g. Clark 1975). At first sight, we might therefore expect dynamical
creation to dominate over destruction also for CVs. However, dynamical
encounter cross-sections scale with mass, so one cannot simply assume
that the same enhancement factor will apply to CVs and LMXBs.

Until recently, estimates of CV formation rates in GCs tended to focus
on one channel at a time. For example, Davies (1997) noted that 
CV progenitors might be destroyed in dense cluster cores, but would
likely survive in the outskirts. His estimate for the number of such
``primordial'' CVs in a given single massive cluster was $N_{CV,p}
\sim 100$, not too different from our naive estimate that ignored
dynamics entirely. The role of two-body encounters in producing GC
CVs was explored by di Stefano \& Rappaport (1994). They found that
the number of CVs produced via the ``tidal capture'' (Fabian, Pringle
\& Rees 1975) of a MS star by a WD might also be $N_{CV,tc} \sim
100$. Finally, Davies (1995) considered 3-body interactions as a
source of CVs, such as exchange encounters in which a WD is exchanged
into a pre-existing MS-MS binary. He found that this channel again
might contribute  $N_{CV,3b} \sim 100$ CVs.

\begin{figure*}[t]
\label{47}
\begin{center}
\resizebox{1.0\hsize}{!}{\includegraphics[clip=true]{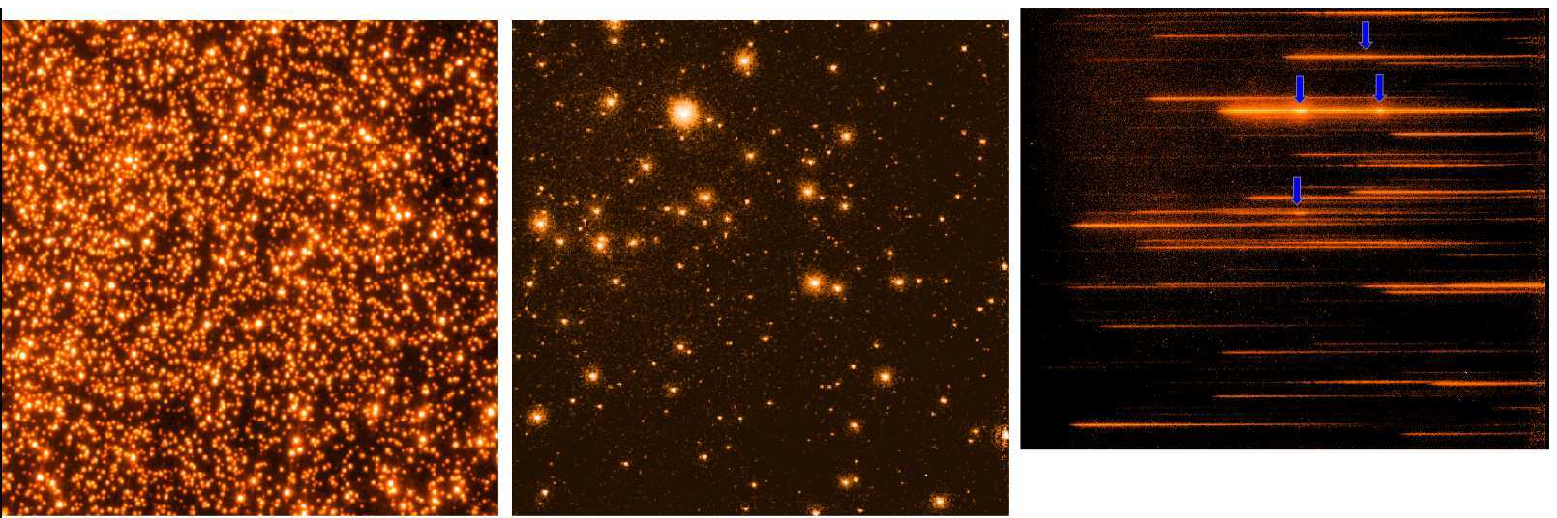}}
\end{center}
\caption{
\footnotesize
{\em Left panel:} A U-band image of part of the core of 47 Tuc. 
{\em Middle panel:} The same region observed in the
far-ultraviolet. {\em Right panel:} A 2-D spectral image of roughly
the same region again, obtained via slitless FUV spectroscopy. Each
horizontal 
trail is the spectrum of a FUV source. Three CVs that exhibit emission
line are marked (see Figure~\ref{spec}). All images span roughly
25\arcsec $\times$ 25\arcsec, 
were obtained with HST and have been described and analyzed fully in
Knigge et al. (2002, 2003, 2008).}
\end{figure*}

\begin{figure}[t]
\resizebox{\hsize}{!}{\includegraphics[clip=true]{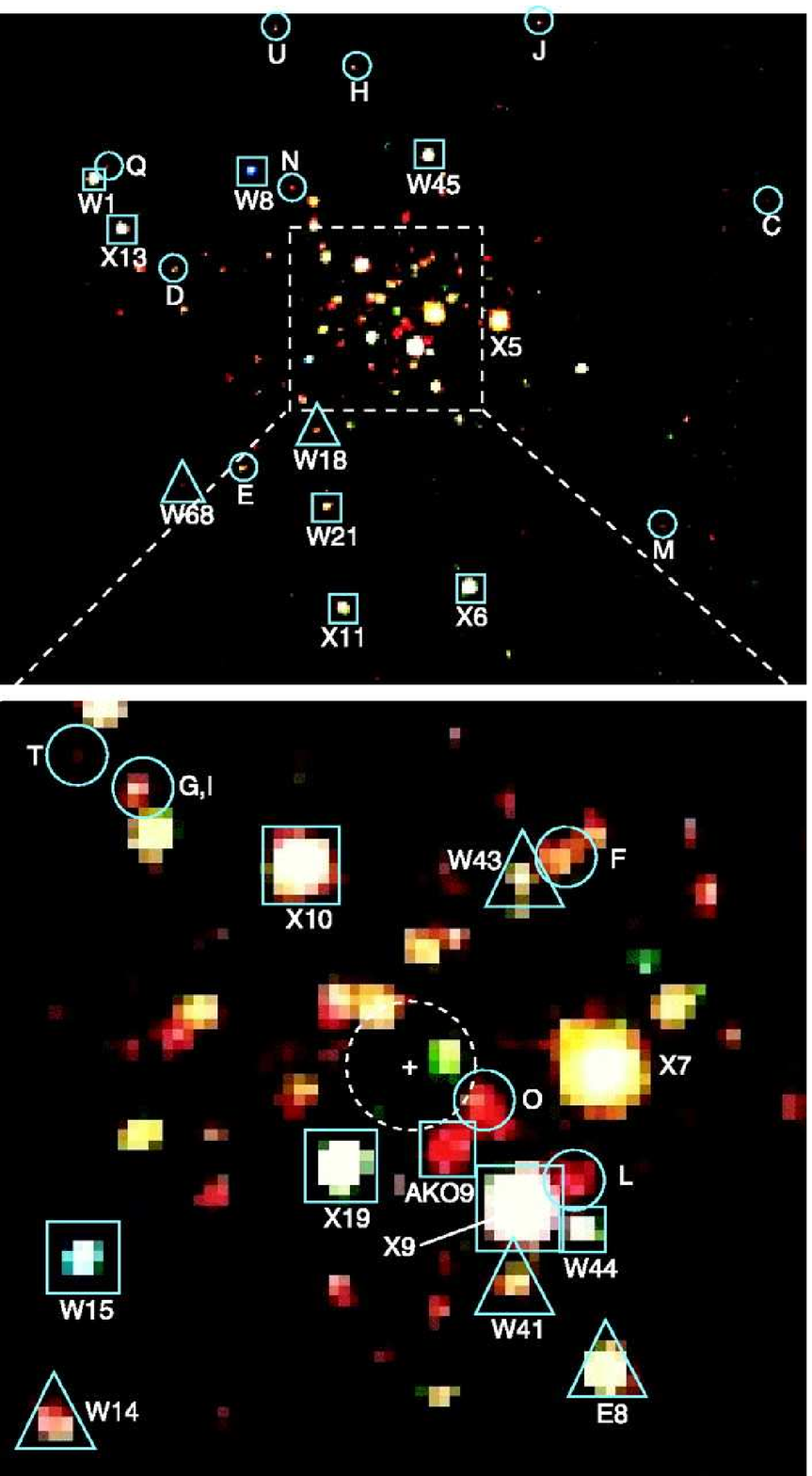}}
\label{xray}
\caption{
\footnotesize
X-ray color image obtained with Chandra of the central 2\arcmin
$\times$ 2.5\arcmin of 47 Tuc. The zoom-in on the core in the bottom
panel spans 35\arcsec square and is thus comparable in size to the
optical and FUV images in Figure~2. CVs are marked with
squares. Figure reproduced by permission from Grindlay et
al. (2001), to which the reader is referred to for details.}
\end{figure}

In recent years, our ability to model GC dynamics and their connection
to binary evolution has improved significantly. We therefore now
have more comprehensive predictions for the number and type of CVs we
might expect to find in GCs (Ivanova et al. 2006). 
Figure~\ref{ivanova} shows the
``flowchart'' of the various CV production channels in a particular GC
model -- real life is clearly more complex than back-of-the-envelope
single-channel estimates. Despite this, the predicted numbers --
$N_{CV,GC} \sim 200$ for a typical massive cluster -- are not far off
the earlier estimates and indicate a mild, $\sim \times 2$ enhancement
of CV numbers over the Galactic field. Nevertheless, the {\em
properties} of the CVs in these models can be quite strange. For
exmaple, the survival rate of primordial CVs is only about 25\%, so
the majority of GC CVs are dynamically formed. Also, a full 60\% of GC
CVs did not form via a common envelope phase, and 50\% formed via some
form of binary encounter.

\section{Detecting CVs in GCs}

So how should we go about detecting the hundreds of CVs that are
predicted to lurk in any given massive GC? The answer is pretty much
in the same way we find them in the rest of the galaxy: via their
variability, their emission lines, their excessively blue colour
and/or their X-ray emission. The GC setting does not exactly make our
life easier though, since (i) GC CVs are typically $10-100$ times more
distant and hence $100 - 10,000$ times fainter than nearby field CVs,
and (ii) GCs are crowded places, with surface densities that can
easily reach or exceed 10 stars per square arcsecond. Nevertheless,
all four of these methods have been used to locate CVs in GCs, so let
us briefly look at each of them in turn.

\subsection{Variabilty}
\label{var}

CVs exhibit many different types of variability on all sorts of time
scales, from flickering and oscillations on time scales of seconds, to
nova eruptions that are thought to recur every $\sim 10^4 {\rm
yrs}$. However, the most obvious type of CV variability to exploit for
locating CVs in GCs is that related to dwarf nova (DN)
outbursts. DNe typically brighten by >3 magnitudes during eruption and 
standard CV evolution theory predicts that {\em most} CVs should be
DNe (e.g. Knigge, Baraffe \& Patterson 2011). So multi-epoch imaging
would 
seem to be a good way to search for CVs in GCs, although we need to
know the characteristic duty cycles of DNe in order to interpret such
observations -- a critical point to which we will return. 

The first concerted effort to use this strategy was undertaken by
Shara et al. (1996), who analysed 12 epochs of HST observations GC 47
Tuc (although only 3 epochs covered more than a fraction of the
cluster core). The recovered only a single -- previously known -- DN and
therefore concluded that {\em ``there are probably no more than three DNe
in the core of 47 Tuc, in significant disagreement with the standard
model of tidal capture, unless the properties of DNe in globulars
differ (e.g. in outburst frequency) from those in the field.''}
Several other DNe in GCs have been discovered since then, including 
two in M80 (Shara, Hinkley \& Zurek 2005), two in NGC 6397 (Shara et
al. 2005), at least one more in 47 Tuc (Knigge et 
al. 2002, 2003); one in M22 (Hourihane et al. 2011) and one in M13
(Servillat et al. 2011). However, the most recent {\em systematic}
study (Pietrukowicz et al. 2008) found {\em ``only 12 confirmed DNe in
a substantial fraction of all Galactic GCs''} and thus concluded that
``the results of our extensive survey provide new evidence...that
ordinary DNe are indeed very rare in GCs''. Taken at face value, these
conclusions are obviously in serious conflict with theory, which
predicts hundreds of CVs in a single massive cluster, most of which
should be DNe. We will return to this apparent conflict in
Section~\ref{dearth}.

\subsection{Emission Lines}

The second obvious way of finding CVs in GCs is via their emission
lines. In practice, such searches have usually been implemented as
narrow-band H$\alpha$ imaging surveys, with crowding still
necessitating the use of HST (e.g. Bailyn et al. 1996; Carson, Cool \&
Grindlay 2000). Typical CV hauls with this method have been a handful of
CV candidates per cluster, i.e. still a long way from the $\sim 200$
predicted. However, an obvious question is whether such searches
reached deep enough uncover the dominant (faint) part of the CV
population. We will come back to this question also; for the moment,
we will merely note that the absolute magnitudes of the CVs recovered
were typically around $M_V \simeq 7-8$. 

\subsection{Blue Colour}

The third way of finding CVs (or at least CV candidates) in GCs is via
their colours. The luminosity of most CVs is
dominated by accretion power, and most of the energy released by
accretion is generated close to the W. These regions are much hotter
than the most massive MS stars in an old GC, so CVs are exceptionally
blue compared to normal stars in a GC. This can already be exploited
in the optical region, where CVs should stand out in, for example, U-B
colour-magnitude diagrams (e.g. Cool et al. 1998), but is also a good
reason to search for GC CVs in the far-ultraviolet (FUV) band with HST
(e.g. Knigge et al. 2002; Dieball et al. 2005, 2007, 2010). The latter
offers the nice additional advantage that crowding ceases to be a problem
entirely, since most ordinary MS stars are effectively invisible in
the FUV (Figure~\ref{47}). On the downside, the field of view of HST's
FUV detectors is very small, and often covers only (part of) the
cluster core. Such searches typically yielded about 1-50 CV candidates
per cluster, but, as before, one needs to keep their limited
completeness in mind.

\subsection{X-rays}

The fourth and final CV detection route we will consider is X-ray
emission. In principle, X-rays are a great way of searching for
compact objects, but, until recently, X-ray observatories suffered
from a combination of poor spatial resolution and limited
sensitivity. Heroic efforts to survey GCs were made nonetheless
(e.g. Hasinger, Johnston \& Verbunt 1994), but even the identification
of individual 
sources -- especially in the critical cluster cores -- was often
impossible. 

However, the current generation of X-ray telescopes -- XMM and,
especially, Chandra -- has completely revolutionized this
field. Chandra, for example, offers excellent sensitivity, a $15\arcmin$
radius field of view and a spatial resolution of $\simeq 1\arcsec$ --
characteristics that are perfectly matched to the requirements of GC
surveys. And, sure enough, the very first survey of a GC with Chandra
(Grindlay et al. 2001; Figure~\ref{xray}) immediately revealed a population of $\simeq
30$ CVs in 47 Tuc, a considerably larger number than had been known
before. Since then, $\simeq 60$ clusters have been surveyed with
Chandra down to $L_x = 10^{31} {\rm erg s}^{-1}$ (Pooley 2010). For
quite a few of these, the limiting $L_x$ is actually considerably
deeper than this, though very few have been searched down to $L_x \leq 10^{30}
{\rm erg s}^{-1}$. These searches typically uncovered 10-100 CV
candidates in any given high-density, massive cluster. This represents
a significant fraction of the predicted population, but is still some
way below the theoretical predictions.

\section{Confirming CVs in GCs}
\label{conf}

\begin{figure}[t]
\label{spec}
\resizebox{\hsize}{!}{\includegraphics[clip=true]{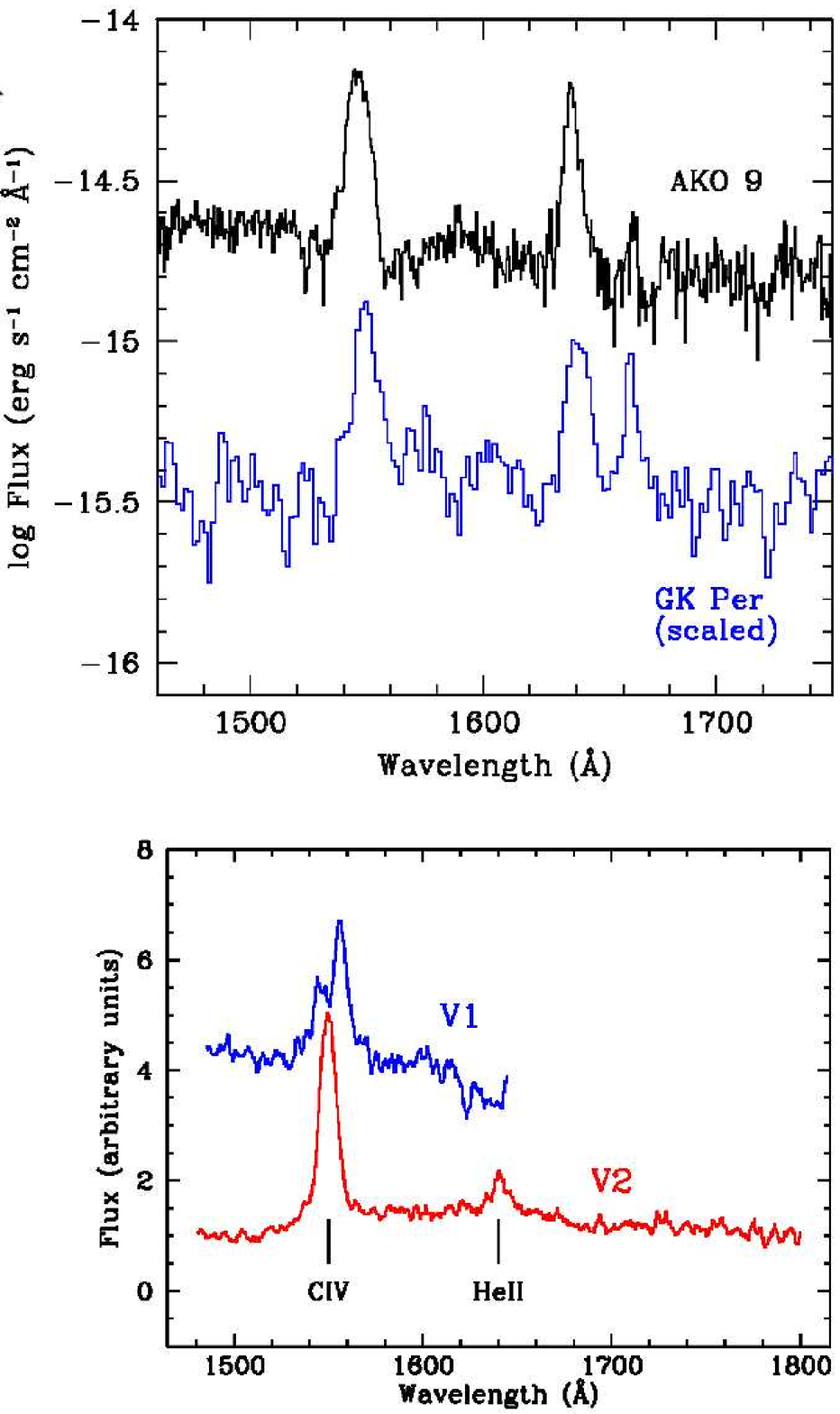}}
\caption{
\footnotesize
FUV spectra for 3 CVs extracted from the 2-D spectral image in the
right panel of Figure~\ref{47}. The top panel shows the spectrum of the
brightest source (AKO 9), compared to a scaled version of the
long-period field CV GK Per. The bottom panel shows the other two
objects highlighted with arrows in Figure~\ref{47}. All three CVs clearly
show emission associated with C~{\sc iv} 1550 \AA; AKO~9 and V2
additionally display He~{\sc ii} 1640 \AA. The top panel is reproduced
from Knigge et al. (2003).}
\end{figure}

An object uncovered by any one of the methods described in the
previous section is obviously just a CV {\em candidate} whose status
as a genuine CV requires confirmation. As in the field, {\em
spectroscopic} confirmation tends to define the 
gold standard  here. However, the faintness of CVs in GCs, coupled
with the crowded 
nature of the GC environment, make this a difficult proposition. In
the optical band, ground-based spectroscopy is only possible in the
outskirts of clusters, and long-slit spectroscopy with HST is an
expensive way of confirming individual CVs. A sneaky and efficient way
around this problem is to exploit the absence of crowding in the FUV 
by carrying out {\em slitless}, multi-object FUV spectroscopy 
(Knigge et al. 2003, 2008; Figure~\ref{47}). In 47 Tuc, this allowed us to
confirm 3 CVs simultaneously, via the detection of the 
C~{\sc iv} 1550~\AA emission line in objects already
displaying a clear FUV excess (Figure~\ref{spec}).

\begin{table*}[t]
\caption{Spectroscopically confirmed GC CVs}
\begin{center}
\begin{tabular}{lccccccc}
\hline
\\
Cluster & Object & Method & Location & Reference \\ \hline
\\
NGC 6397 & CV 1  & HST, opt & core & (1) \\
NGC 6397 & CV 2  & HST, opt & core & (1) \\
NGC 6397 & CV 3  & HST, opt & core & (1) \\
NGC 6397 & CV 4  & HST, opt & core & (2)  \\
NGC 6624 & CV    & HST, FUV & core & (3) \\ 
47 Tuc  & AKO 9 & HST, FUV, slitless & core & (4)\\
47 Tuc  & V1    & HST, FUV, slitless & core & (5)\\
47 Tuc  & V2    & HST, FUV, slitless & core & (5)\\ 
M 5      & V101  & ground-based, opt & outskirts & (6) \\ \hline
\end{tabular}
\end{center}
\footnotesize{\em References:} (1) Grindlay et al. 1995; (2) Edmonds
et al. 1999; (3) Deutsch et al. 1999; (4) Knigge et al.(2003); (5)
Knigge et al. 2008; (6) Margon, Downes \& Gunn 1981
\end{table*}

However, on the whole, the gold standard has been rarely achieved to
date. In fact, only 6 studies have so far succeeded in
spectroscopically confirming one or more new GC CVs, and the combined
total adds up to {\em only 9 spectroscopically confirmed CVs in just 4
GCs} (Table~1). 

So what is the silver standard? Usually, it is confirmation by
multiple methods. For example, an object exhibiting a 3-magnitude
outburst that is also an X-ray source and abnormally blue in the
optical is pretty much guaranteed to be a DN-type CV. In practice,
part of such confirmations can come almost for free. For example, Cohn
et al. (2010) use $H\alpha$ imaging with HST to confirm Chandra X-ray
sources in NGC~6397 as CVs. But since broad-band exposures are much
shorter than $H\alpha$, they can be obtained at essentially no extra
cost. Moreover, the $H\alpha$ imaging can be split into short
sub-exposures, allowing a search for short time-scale variability
(e.g. flickering). Thus Cohn et al. (2010) are able to confirm X-ray
CV candidates very effectively through a combination of $H\alpha$
excess, blue colours and variability. 

Similar studies -- though not usually based on the same
combination of methods -- have been carried out in a few GCs. Among
these, that by Edmonds et al. (2003ab) deserves particular comment,
since it also yielded orbital periods for several X-ray-detected CVs
in 47 Tuc (see Section~\ref{porb}). However, there is a significant downside
to the use of all of these different methods for confirming CVs: it
leaves us with an overall sample of CVs that suffers from strong,
multiple and ill-defined selection effects. 

\section{Population Properties}

Having discussed how CVs in GCs are found and confirmed, let us now
consider what we have learned about their global properties. We need
to be clear from the outset, however, that the statistical
properties of observational CV samples in GCs are severely distorted
by the selection biases we have just discussed. 

\begin{table*}
\caption{GC CVs with Orbital Periods}
\begin{center}
\begin{tabular}{lccccccc}
\hline
\\
Cluster & Object & $P_{orb}$ (hr) & Type & Method & References & Comments \\ \hline
\\
47 Tuc & AKO 9 & 26.6 & DN & Ecl,Opt & 1,2,3 & similar to GK Per \\
47 Tuc & V3,W27?& 4.7 & polar? & Var,X-ray & 4,5 & qLMXB? \\
47 Tuc & W1     & 5.8 &        & Var,Opt & 6,7 & $P_{orb} = 5.8/2$ hr?\\
47 Tuc & W2     & 6.3 &        & Var,Opt & 6,7 & no opt signal \\
47 Tuc & W8     & 2.9 &        & Ecl,Opt & 6,7 & \\
47 Tuc & W15    & 4.2 &        & Ecl,Opt & 6,7 & \\
47 Tuc & W21    & 1.7 &        & Var,Opt & 6,7 & \\
47 Tuc & W71?   & 2.4 &        & Var,Opt & 6,7 & really a CV?\\
47 Tuc & W120   & 5.3?&       & Var,Opt & 6,7 & \\
NGC 6397 & CV1 & 11.3 & & Ecl,Opt & 8,9 & \\
NGC 6397 & CV6 & 5.6 & & Ecl,Opt & 8,9 & \\
M 5      & V101 & 5.8 & DN & RV,Opt & 10 \\ 
NGC 6752 & V1   & 5.1 &    & Var,Opt & 11 \\
NGC 6752 & V2   & 3.7? &    & Var,Opt & 11 \\
M 22     & CV 2 & 2.1 & DN  & SH,Opt & 12 & $P_{orb}$ est. from superhumps\\
\hline
\end{tabular}
\end{center}
\footnotesize{\em References:} (1) Edmonds et al. 1996; (2) Albrow et
  al. 2001; (3) Knigge et al. 2003; (4) Shara et al. 1996; (5)
  Heinke et al. 2005; (6) Edmonds et al. 2003a; (7) Edmonds et
  al. 2003b; (8) Kaluzny \& Thompson 2003; (9) Kaluzny et al. 2006; (10)
  Neill et al. 2002; (11) Bailyn et al. 1996; (12) Pietrukowicz et
  al. 2005.\\
Notation for method is var = photometric variability; ecl = eclipses;
RV = radial velocity; SH = superhumps
\end{table*}

\subsection{Orbital Periods}
\label{porb}

The orbital period distribution of CVs has arguably been the most
powerful observational tool for studying CV formation and evolution in
the Galactic field. The famous ``period gap'' -- the dearth of CVs
with $2 {\rm hr} \ltappeq P_{orb} \ltappeq 3 {\rm hr}$ -- as well as
the existence of a minimum period $P_{min} \simeq 80 {\rm min}$ have
been the key properties that led to the development of the ``standard
model'' for CV evolution. In this model, magnetic braking moves CVs
from long to short $P_{orb}$ above the period gap, while gravitational
radiation drives the evolution below the gap (see Knigge, Baraffe \&
Patterson 2011 for a comprehensive overview of CV evolution). 

\begin{figure}[t]
\label{periods}
\resizebox{\hsize}{!}{\includegraphics[clip=true]{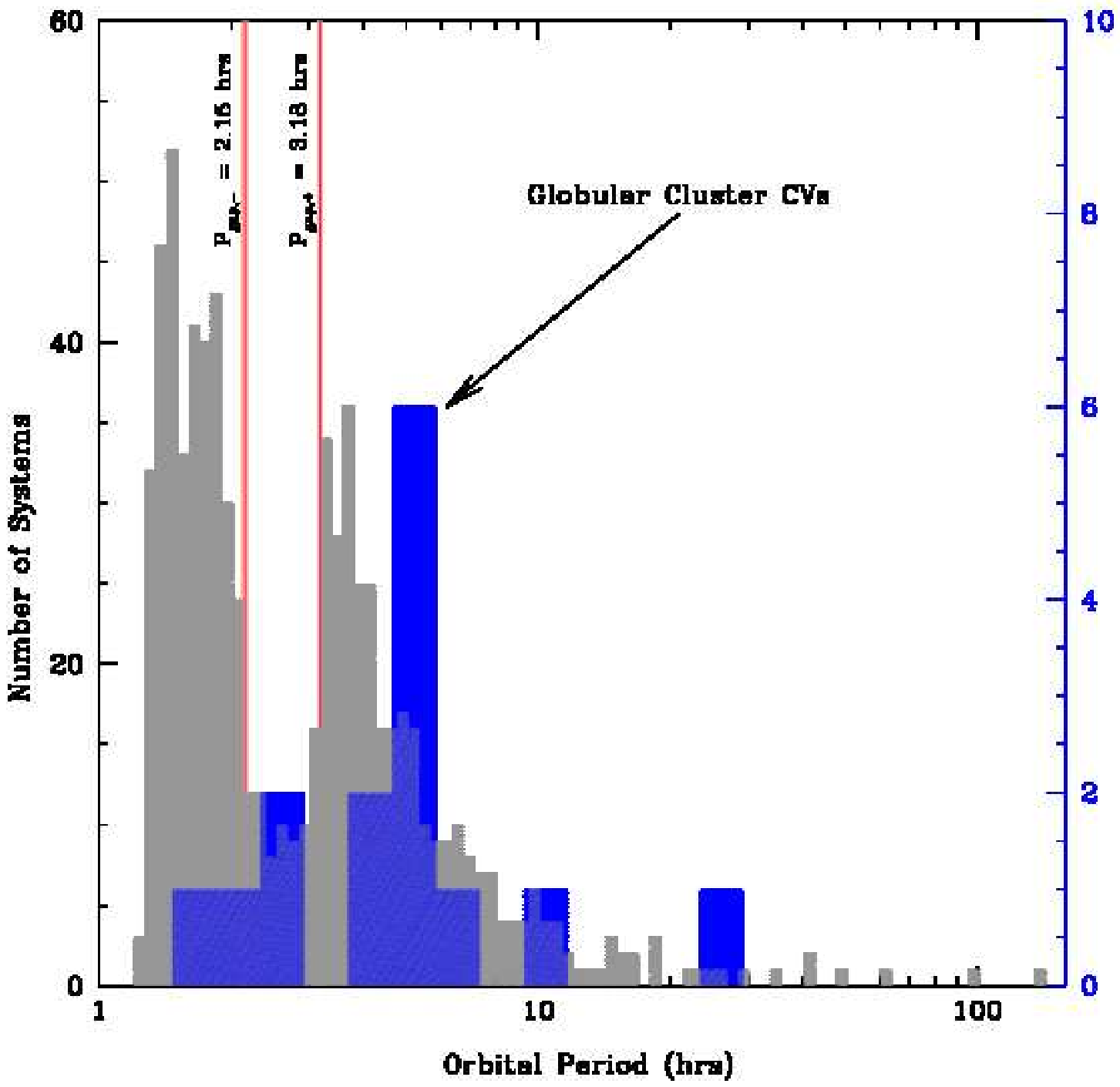}}
\caption{
\footnotesize
The observed period distributions of field and GC CVs. The grey
histogram shows the distribution for field CVs from a recent edition
of the Ritter \& Kolb (2003) catalogue. The blue histogram is the
(sparse) GC period distribution collated from the literature. The
location of the period gap as determined by Knigge (2006) is
indicated by the vertical lines.}
\end{figure}

Unfortunately, orbital periods are known for only 15 CVs in 6 GCs (see
Table~2). Nine of these CVs are located in 47 Tuc, mostly due to the
excellent work by Edmonds et al. (2003ab). Such a small and biased sample
is not particularly well suited to statistical analyses, but
let us throw caution to the wind and compare the $P_{orb}$
distribution of CVs in GCs to that observed in the Galactic field
(Figure~\ref{periods}). This comparison reveals that the existing GC CV sample
shows no sign of the period gap, and indeed only 2 GC CVs
($\simeq 13\%$ of the sample) have short periods below the gap. Both
of these properties stand in stark contrast to the field population.

The obvious and naive interpretation of this comparison is that the
properties of GC CVs are fundamentally different from those of CVs in
the Galactic field. This is entirely plausible, given that most GC CVs
are expected to have been produced or affected by dynamical encounters
(see Section~2). Indeed, some theoretically predicted period
distributions for GC CVs 
look superficially quite similar to the observed distributions (Shara
\& Hurley 2006; Ivanova et al. 2006; but note that in both cases, the
predicted distributions are still affected by technical/computational
issues).

All of this sounds quite promising, but we should remember a hard
lesson learned from field CVs. Here, we have known for a
long time that predicted and observed period distributions tend to
disagree quite badly. For example, theoretical binary population
synthesis models 
predict that a full $\simeq 99\%$ of Galactic CVs in the field should
be short-period systems below the gap (e.g. Kolb 1993) and that there
should be a pile-up of systems at the minimum period (e.g. Kolb \&
Baraffe 1999). Until recently, neither of these predictions appeared
to be consistent 
with the properties of observed samples, and there was much debate
about whether this meant that the models were wrong or whether the
apparent disagreements were all caused by selection effects. This
debate is still not fully settled, but what {\em has} become 
quite clear is that selection effects are extremely
important and must be accounted for in any meaningful comparison of
theory and observation (e.g. Pretorius, Knigge \& Kolb 2007). In fact,
the period spike at $P_{min}$ appears to have been discovered now,
thanks entirely to the construction of a CV sample that is deep enough
to actually detect these very faint CVs in significant numbers
(G\"{a}nsicke et al. 2009). 

With this cautionary tale in mind, I believe it is too early to draw
dramatic conclusions from plots like Figure~\ref{periods}. What is clear
though is that we desperately need a much larger, and ideally more
cleanly selected, sample of GC CVs with orbital periods.

\subsection{Scaling with Collision Rate: Do Dynamics Matter?}

There is another way to test if dynamical effects are producing many
or most CVs in GCs, which does not require any binary parameter
determinations for individual CVs at all. The idea is simple: if most
CVs are formed dynamically, then their number in a given cluster
should scale with the rate at which dynamical encounters take place in
that cluster. This so-called ``collision rate'' or ``encounter
frequency'' is a function only of cluster (or, rather, cluster core)
parameters. The existence of such a scaling for LMXBs in GCs was
conclusively established by Pooley et al. (2003), confirming earlier
work dating back to at least Verbunt \& Hut (1987). 

\begin{figure}[t]
\label{coll}
\resizebox{\hsize}{!}{\includegraphics[clip=true]{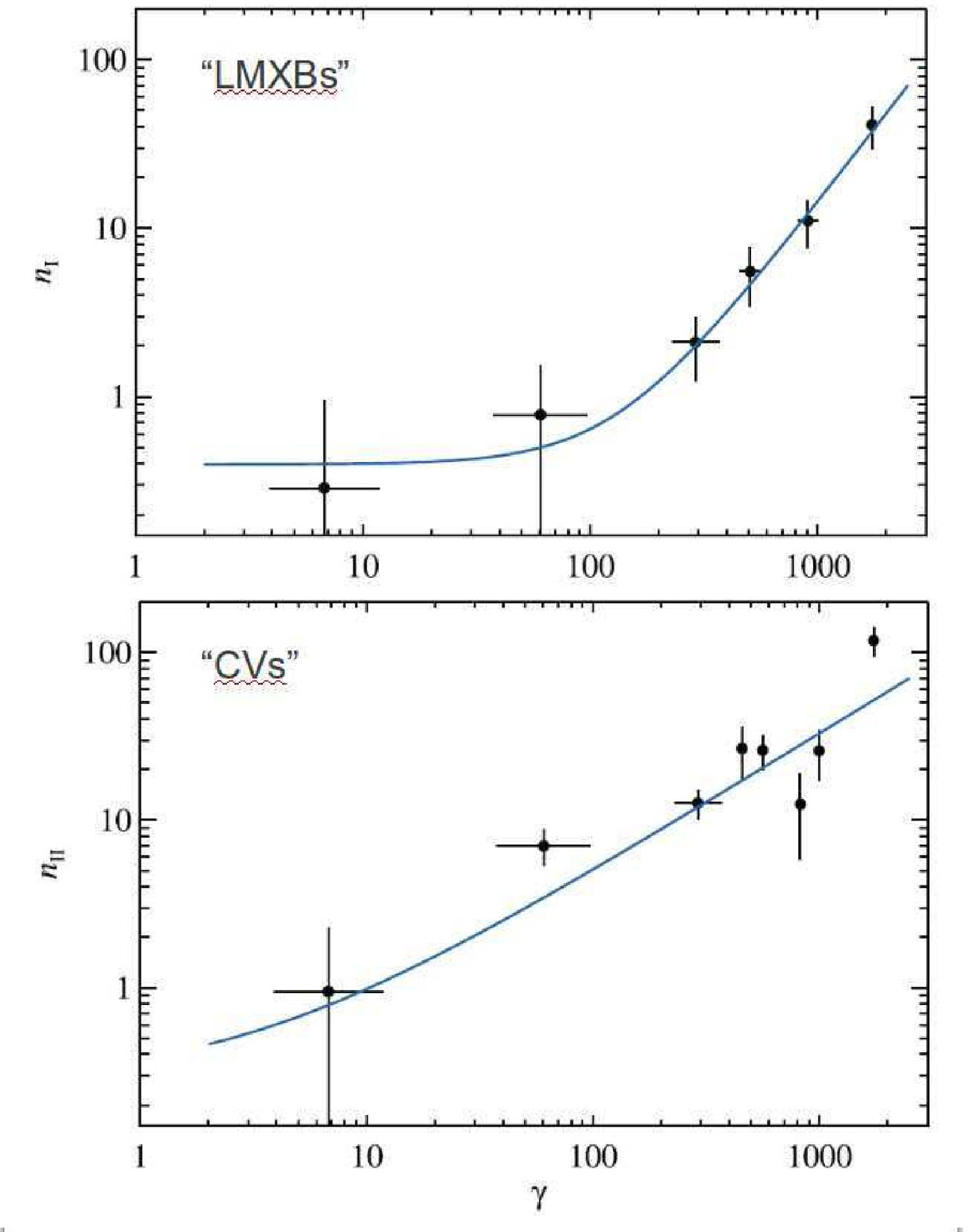}}
\caption{
\footnotesize
The specific number of probable LMXBs (top panel) and
CVs (bottom panel) in GCs as a function of the specific stellar encounter
rate in the cluster. The data were binned for plotting. In both 
cases, there is a clear correlation, which strongly suggests that both of
these compact binary populations are dominated by dynamically formed
systems in most GCs. Figure reproduced by permission from Pooley \&
Hut (2006).}
\end{figure}

Figure~\ref{coll}, from Pooley \& Hut (2006), shows how both LMXB and CV
numbers in GCs scale with collision rate. Actually, what we have
called ``CVs'' and ``LMXBs'' here are really X-ray sources with $L_x >
4 \times 10^{31} {\rm erg~s}^{-1}$, divided into these two categories
on the basis of an empirical cut in the X-ray colour-magnitude
diagram, but Pooley \& Hut (2006) argue convincingly that these
populations will be dominated by CVs and LMXBs,
respectively. Moreover, the quantities actually plotted in Figure~\ref{coll}
the {\em specific} number of objects (i.e. CVs/LMXBs per
$10^6 {\rm M_{\odot}}$) against the {\em 
specific} encounter rate (a measure of the rate at which an individual
star undergoes encounters). Ordinary stellar populations should trace
a straight line across such a plot, while entirely dynamically-formed
ones might be expected to follow a straight line with slope unity (but
note that the plots are logarithmic and also that this prediction
ignores subtleties like that non-constancy of the collision rate in a
given cluster). Figure~\ref{coll} clearly shows that there is a scaling of CV
numbers with encounter frequency, which strongly supports the idea
that most CVs in GCs are indeed dynamically formed. 

\subsection{X-ray and Optical Properties}
\label{xopt}

An in-depth study of the X-ray and optical properties of GC CVs was
carried out by Edmonds et al. (2003ab), who specifically looked at the
X-ray selected CV sample in 47 Tuc. One key set of tests they performed
was to compare the 47 Tuc CVs to a sample of X-ray-detected field CVs
(taken from Verbunt et al. 1997) in terms of the optical magnitudes,
X-ray luminosity and X-ray-to-optical ratio. Their maindings were as
follows: (i) optically, GC CVs are rather faint and most
similar to dwarf novae; (ii) however, their X-ray luminosities are
higher than those of non-magnetic field CVs, and most resemble those
of intermediate polars (IPs); (iii) unsurprisingly then, their
X-ray-to-optical ratios are abnormally high and not really consistent
with any field CV sub-population.

So does this mean that most GC CVs are somehow physically different
from most field CVs? Perhaps. Edmonds at al (2003ab) themselves
suggest that their findings may point to GC CVs being peferentially
magnetic, low-$\dot{M}_{acc}$ systems. However, we should treat such 
comparisons of field and GC CV samples with a great deal of caution,
since selection effects could easily lead to misjudgments. For
example, the GC CV sample studied by Edmonds et al. (2003ab) was
entirely X-ray selected (which would always favour the detection of
magnetic CVs). Moreover, their field CV sample is a set of objects
detected by ROSAT that were {\em known to be CVs before ROSAT was
launched and had known orbital periods} (Verbunt et al. 1997). As
such, they form a totally heterogenous sample selected via multiple
methods. It would be interesting and worthwhile to repeat the field vs
GC CV comparison with an X-ray-selected field CV sample. My own
feeling is that the high X-ray-to-optical ratio of the 47 Tuc CV
sample is an important clue, but that it is far from clear what it 
points to.

\subsection{Are Most GC CVs Magnetic?}

As noted above, Edmonds et al. (2003ab) argued that the properties of
the X-ray-selected CV sample in the famous GC 47 Tuc pointed towards a
pre-dominance of magnetic CVs in GCs. This idea actually goes back to
at least Grindlay et al. (1995), who obtained spectra of 3 CVs in
NGC~6397 with HST (see Section~\ref{conf} and Table~1) and noted the
existence of He~{\sc ii} emission in all of them. This line points to
the presence of a strong EUV source and is commonly -- but by no means
exclusively -- seen in magnetic CVs, where the polar accretion caps on
the WD provide such a source. However, non-magnetic nova-like CVs, in
particular, also often show this features.

On its own, the presence of He~{\sc ii} lines in a few GC CVs is of
course rather weak evidence that most GC CVs are
magnetic. Nevertheless, this idea has gained considerable traction
over the years. The main reason for this is the hope that it may 
explain some of the apparent abnormalities of the GC CV 
population, such as the strange X-ray and optical characteristics
(Edmonds et al. 2003ab) and the low discovery rate of DNe in GCs 
(e.g. Shara et al. 1996; Dobrotka, Lasota \& Menou 2006; Pietrukowicz et
al. 2008 -- see Section~\ref{var}). 

From a theoretical point of view, a preference for magnetic CVs in
GCs seems unlikely at first sight, but there actually is a possible mechanism
that could explain this. As first pointed out by Ivanova et
al. (2006), magnetic WDs (in the field) tend to be relatively massive
(e.g. Vennes 1999), and since the cross-section for dynamical
encounters scales with mass in GCs, this might make magnetic WDs more
likely to form CVs dynamically.

So should we accept that GC CVs are preferentially magnetic? Once
again, the evidence needs to be evaluated with considerable
caution. As discussed in Section~\ref{xopt}, the X-ray and optical
properties of the 47 Tuc CVs do {\em not} point cleanly to a sample
dominated by magnetic systems. In fact, the only property that seemed
to match well to that of field CVs (subject to the usual caveats
regarding selection effects), was the X-ray luminosity
distribution. More recently, Heinke et al. (2008) have shown that the
global X-ray {\em spectral} properties of GC CVs are best matched if
the propertion of magnetic CVs is ``only'' $\simeq 40\%$. This number
is still larger than the fraction of magnetic CVs among all known
field CVs, which is $\simeq 10\% - 20\%$ (Ritter \& Kolb 2003; Downes
et al. 2005). However, as already noted by Heinke et al. (2008), one
needs to keep in mind that the GC CV sample is X-ray selected, which
would favour the detection of magnetic systems (since these are known
to be X-ray bright). In fact, a full 64\% (29/45) of the CVs found in
the Rosat Bright Survey (Schwope et al. 2002; also see Pretorius \&
Knigge 2011) are magnetic, although this comparison is also unfair, 
since that survey is flux-limited and not particularly deep (so the
volume sampled for X-ray bright magnetic CVs is much larger than that
for non-magnetic systems). X-ray surveys of GC that go deep enough to
detect {\em all} non-magnetic CVs would, of course, be unbiased, but
few, if any, of these have been carried out. 
\footnote{In line with this, Servillat et al. (2008) conclude that
their joint X-ray and UV survey of NGC 2808 does not allow them to
"confirm or rule out a possible excess of IPs'' in this cluster.}
This still leaves the apparent dearth of DN outbursts in GCs as a
possible piece of circumstantial evidence in favour of
magnetically-dominated CV population in GCs, so let us take a closer
look at this.

\subsection{Are DNe Abnormally Rare in GCs?}
\label{dearth}

As already noted in Section~\ref{var}, both Shara et al. (1996) and
Pietrukowicz et al. (2008) concluded that DNe are rarer than
expected in GCs if (a) there are really hundreds of CVs in GCs,
and (b) cluster DNe have similar properties to known field
DNe. 

Obviously, what is actually observed in these multi-epoch surveys is
a particular number of eruptions over a certain time scale. However,
in order to interpret these observations, we need to know the
completeness of the survey, i.e. the fraction of DNe it is expected to
have recovered. Both Shara et al. (1996) and Pietrukowicz et
al. (2008) took pains to estimate this via Monte Carlo simulations, 
using well-known field DNe light curves as templates.

The single most important light curve parameter for determining the
completeness of such a survey is the characteristic duty cycle,
$f_{on}$, of DNe,  i.e. the fraction of time that a typical DN spends
in outburst. Specifically, the completeness of a DN survey that
consists of N independent epochs is $\epsilon = 1-(1-f_{on})^N$ This
is a key point. Both  Shara et al. (1996) and Pietrukowicz et
al. (2008) based their efficiency estimates on the properties of
well-known field DNe. But {\em that} sample is extremely
unlikely to be representative of the underlying CV population even in
the Galactic field. More specifically, these systems were probably
discovered and well-observed precisely because they are bright, {\em
frequently erupting} long-period systems, i.e. because they have a
high duty cycle.

The duty cycles of the DNe templates used by Shara et al. (1996) range
from roughly 15\% to over 95\% with an average of about 
50\%, while the two control DNe used by Pietrukiwicz et al. (2008)
spend roughly 10\% and 30\% of their lives in eruption. Shara et
al. (1996) also presented an analytic efficiency estimate based on an
assumed duty cycle of 15\%. Given these numbers, let us take $\simeq$
20\% as a rough estimate of the characteristic DN duty cycle assumed
in both studies. 

How does this number compare to the duty cycle of the most common DNe 
among the intrinsic CV population. As already noted above, theory
predicts that the vast majority of CVs should be short-period systems,
and indeed most of them should be ultra-faint CVs at or near the
minimum period. (Note that only 5 of the 21 DNe on which the Shara et
al. templates were built are short-period systems, and neither of the
two control DNe used by Pietrukiwicz et al. are.) These faint,
short-period systems are almost certainly still massively
under-represented in the known field CV sample, so their
characteristic properties are necessarily uncertain. 

The best-known example of such a system is probably WZ Sge, so let us take
a look at its duty cycle. WZ Sge erupts as a DN roughly once every 30
years, and its main eruption lasts about 1-2 months (Patterson et
al. 2002). This amounts to a duty cycle of about 0.4\%! If such CVs
were to dominate the CV population in GCs, the completeness of 
existing GC DNe surveys has been seriously overestimated. For the case
of N = 3 (the number of epochs covering the full core in Shara et
al. 1996), a change from $f_{on} = 0.2$ to $f_{on} = 0.04$ reduces
the completeness by a factor of 40.\footnote{I should note explicitly
here that both Shara et al. (1996) and Pietrukowicz et al. (2008)
fully acknowledge the low completeness of their surveys for WZ Sge-like
systems. What I am stressing here is that such systems might be {\em
expected} to be the dominant CV population.} So even if $\sim 150$ WZ Sge-like
CVs were to lurk in the core region of 47 Tuc, only about 2 should have been
detected. Indeed, Shara et al. actually detected two DNe in their survey
(V2 and V3), although neither is likely to be a WZ Sge-like
object. But even the Poisson probability of seeing zero systems when
just under 2 are expected is a healthy 15\%. Servillat et al. (2011)
carried out a similar calculation -- and arrived at similar
conclusions -- for their variability survey of M13.

All of this may be overly pessimistic (or optimistic, depending on
one's viewpoint). And the completeness of the survey by Pietrukowicz
et al. (2008) for at least some clusters is higher than that achieved
by Shara et al. (1996) for 47 Tuc and Servillat et al. (2011) for
M13. But it is difficult to be sure. The problem 
is that we simply do not know the intrinsic outburst properties
properties of even the {\em field} CV/DNe population well enough. What
is needed is a survey for DNe in GCs that {\em guarantees} the
detection of at least a few WZ-Sge-like systems, if they exist in
large numbers in GCs. This is difficult, but not impossible.
 
\section{Conclusions}

I hope to have shown that GC CVs should be of great interest to
anybody studying the dynamical evolution of clusters, the binary
evolution of CVs or the SN Ia progenitor problem. I also hope I have
made it clear that great strides have been taken in recent years in
finding and understanding GC CV populations, thanks mainly to the
availability of Chandra and HST. Indeed, we are finally discovering
significant numbers of CVs in any given massive GC, though still not
the hundreds that are theoretically predicted to lurk there. I have
also discussed the theoretical and observational hints that field and
GC CV populations may differ systematically, although I feel that the
jury is still out regarding most of these putative differences.

The next big advances in the field are likely to come from one or
both of the following directions. First, if we really wish to test for
the presence of the predicted hundreds of CVs per cluster, we have to
be sensitive to {\em faint} CVs. How deep? Ideally, deep enough to discover
systems like WZ Sge, at $M_{V} \sim 13$ in quiescence. This is hard --
it requires reaching $m_v \sim 27$ or so. But it is not impossible:
Cohn et al. (2010) have essentially done this in NGC~6397 and uncovered
two likely WZ-Sge-like CVs, the first such systems in any GC. An
interesting short-cut might be to place limits on the number of faint
CVs in GCs by considering the {\em integrated} X-ray luminosity of
individually undetected systems (e.g. Haggard, Cool \& Davies
2009). Second, we need to obtain orbital periods for a significant
sample of GC CVs, so that we can start to make meaningful comparisons
to theoretically predicted period distributions.


\section{Postscript: Three Topics Missing From This Review...}

There are three topics that I really wanted to cover in this
review, but then I simply ran out of time and space. In the few lines
I have left here, let me at least mention them. First, {\em novae}
might both clear (Moore \& Bildsten 2011) and enrich 
(Maccarone \& Zurek 2011) GCs; the latter might be key for the
multiple stellar populations recently found in some GCs. At least one
GC nova has been recovered (Dieball et al. 2010). Second, many 
{\em AM CVns} should exist in GCs (Ivanova et al. 2006), yet not a
single one is known. Third, {\em symbiotic stars} in GCs also remain
to be found -- but we may just have discovered the first one (Zurek et
al., in prep.).  




\bibliographystyle{aa}

\end{document}